\documentclass{elsart5p}
\usepackage{amssymb}
\usepackage{graphicx}
\usepackage{bm}

\begin{document}

\begin{frontmatter}

\title{Correlated transport of FQHE quasiparticles in
a double-antidot system}

\author{Dmitri V. Averin and James A. Nesteroff}

\address{Department of Physics and Astronomy, Stony Brook
University, SUNY, Stony Brook, NY 11794-3800 }

\date{\today}

\begin{abstract}
We have calculated the linear conductance associated with
tunneling of individual quasiparticles of primary quantum Hall
liquids with filling factors $\nu =1/(2m+1)$ through a system of
two antidots in series. On-site Coulomb interaction simulates the
Fermi exclusion and makes the quasiparticle dynamics similar to
that of tunneling electrons. The liquid edges serve as the
quasiparticle reservoirs, and also create the dissipation
mechanism for tunneling between the antidots. In the regime of
strong dissipation, the conductance should exhibit resonant peaks
of unusual form and a width proportional to the quasiparticle
interaction energy $U$. In the weakly-damped regime, the shape of
the resonant conductance peaks reflects coherent tunnel coupling
of the antidots. The Luttinger-liquid singularity in the rates of
quasiparticle tunneling to/from the liquid edges manifests itself
as an additional weak resonant structure in the conductance
curves.

\end{abstract}

\end{frontmatter}

\section{Introduction}
\label{sec1}

Quasiparticles of two-dimensional (2D) electron liquids in the
regime of the Fraction Quantum Hall effect (FQHE) have the unusual
properties of fractional charge \cite{b1} and fractional exchange
statistics \cite{b2,b3}. Quantum antidots formed in the 2D
electron system offer a possibility of localizing and controlling
transport of individual quasiparticles \cite{ant1,ant2,ant3}. Such
a control made possible the first direct observation of the
fractional quasiparticle charge in tunneling through an antidot
\cite{ant1}. This observation was later extended to the regime of
ballistic quasiparticle transport \cite{b4,b5}.

The mechanism of quasiparticle localization on antidots relies on
the combined action of the electric and magnetic field and is
microscopically quite different from the corresponding features of
electron localization in quantum dots. Nevertheless, transport
phenomena in antidots are very similar to those associated with
the Coulomb blockade in tunneling of individual electrons
\cite{al}. For instance, in close analogy to the Coulomb-blockade
oscillations of conductance of the quantum dots \cite{dot}, the
antidots exhibit periodic conductance oscillations with each
period corresponding to addition of one quasiparticle to the
antidot \cite{ant1,ant2,ant3}. So far, antidot transport has been
studied both experimentally \cite{ant1,ant2,ant3} and
theoretically \cite{gel} for one antidot. In this work, we develop
a theory of correlated transport of individual quasiparticles
through two antidots. The double-antidot system was discussed
previously \cite{qub} as a qubit, information in which is encoded
by individual quasiparticles. Such qubit is similar to
superconducting charge qubits \cite{sq1,sq2} which are based on
the dynamics of individual Cooper pairs. As in the case of Cooper
pairs \cite{t1,t2}, the transport measurements on the
quasiparticle qubit can be done more easily that direct
measurements of the qubit dynamics. Transport measurements would
constitute the first step towards experimental development of the
FQHE qubits. More generally, understanding the transport
properties of multi-antidot systems, in particular the role of
Coulomb interaction for localization of individual quasiparticles,
and significance of the edge-state decoherence, should also be
important for other, more complicated types of suggested FQHE
qubits \cite{na1,na2} which also require control over individual
quasiparticles.

\section{Model}
\label{sec2}

\begin{figure}
\hspace*{.05in}
\includegraphics[width=3.2in]{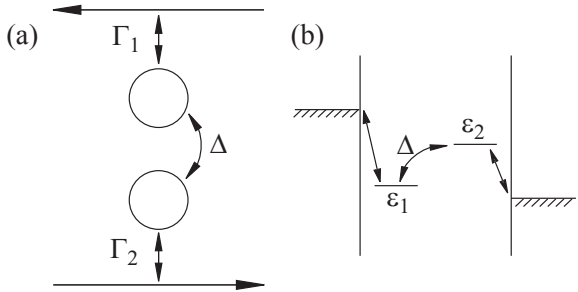}
\setlength{\unitlength}{1.0in}
\caption{Quasiparticle tunneling in the double-antidot system: (a)
the real-space geometry (not to scale) of quasiparticle transfer
between the opposite edges of the FQHE liquid; (b) energy diagram of
the transfer. } \label{f1}
\end{figure}

The system we consider consists of two antidots in series between
the two opposite edges of a primary quantum Hall liquid with the
filling factor $\nu =1/(2m+1)$ (Fig.~\ref{f1}a). The antidots are
tunnel-coupled to each other and to the edges, which play the role
of quasiparticle reservoirs. The quasiparticle current through the
antidots is driven by the transport voltage $V$ applied between
the edges. The focus of this work is on the regime when all
relevant energies are smaller than the energy gap $\Delta^*$ of
the antidots (see below), and the transport can be described
completely in terms of the transfer of individual quasiparticles.
This regime is relevant, e.g., for the operation of this system as
a qubit. The main elements of the model of the double-antidot
system in this case can be outlined as follows.

\subsection{Antidots}

An antidot formed at a point $\zeta$ in a primary quantum Hall
liquid with the filling factor $\nu =1/(2m+1)$ can be described as a
collection of $n$ quasihole excitations created at this point.
Microscopically, the unnormalized wavefunction of this configuration
is \cite{b1}:
\begin{equation}
\psi(\{ z_j \})= \prod_j (z_j-\zeta )^n \psi_m (\{ z_j \}) \, ,
\label{e1} \end{equation} where $\psi_m (\{ z_j \})$ is the
Laughlin's wavefunction of the unperturbed quantum Hall liquid
and, in the standard notations, the antidot position $\zeta$ in
the two-dimensional plane and the electron coordinates $z_j$ are
given in the complex form and normalized to the characteristic
length $\ell=(\hbar/eB)^{1/2}$ in the magnetic field $B$. The
number $n$ of the quasiholes is related to the geometric radius
$R$ of the antidot: $R \simeq \sqrt{2n} \ell$.

In what follows, we make use only of the general qualitative
features of the wavefunction (\ref{e1}). For instance, in agreement
with the typical experimental situation (see, e.g., \cite{ant1}),
we assume that the antidot is relatively large: $n\gg 1$. This
means that addition/removal of individual quasiparticles (here and
below, this term will be used to describe processes with varying
$n$: $n\rightarrow n\pm 1$), does not change the antidot parameters
noticeably. Indeed, the variation of the antidot radius in this
case is $\delta R \propto \ell / \sqrt{n}$ and is small not only
on the scale of $R$, but, more importantly, on the scale of the
magnetic length $\ell$.

The general form of the antidot energy $E_n$ as a function of $n$
is determined by the interplay of Coulomb interaction and an
external potential used to create the antidot. However, for large
$n$, and in some small range of variation of $n$ around the minimum
of $E_n$, one can always approximate this dependence as quadratic.
This defines the characteristic energy gap $\Delta^* \equiv
\partial^2 E_n/\partial n^2$ which gives the energy interval of
variation of the chemical potential $\mu$ of the system between the
successive additions of individual quasiparticles to the antidot.
For the system shown in Fig.~\ref{f1}a, the antidots exchange
quasiparticles with the edges, and $\mu$ is defined by the
edge chemical potential. In the situation of the antidot, when all
the energies are dominated by the Coulomb repulsion, the energy
gap $\Delta^*$ for changing the number of quasiparticles is
approximately related to the energy gap for the antidot
excitations at fixed $n$: $\Delta^* \simeq \hbar u /2 \pi R $,
where $u$ is the velocity of the excitations encircling the
antidot. In general, e.g. in the case of quantum dots, the two
types of energy gaps can be very different.

We assume that the gap $\Delta^*$ is sufficiently large for both
antidots of the double-antidot system, so that in the reasonably
large range of variation of $\mu$ both antidots are characterized
by some well-defined numbers $n_l$, $l=1,2$, of the quasiparticles.
In this regime, the non-vanishing conductance of the double-antidot
system requires that $\mu$ is close to resonances in the both
antidotes. At resonance, $E_{n_l}\simeq E_{n_l+1}$, and each antidot
can in principle be in one of two states which differ by the
presence/absence of one ``extra'' quasiparticle. The resulting
four double-antidot states are relevant for the quasiparticle
transport. We will use the notation for these states that
gives the number of extra quasiparticles on each antidot:
\begin{equation}
|ij\rangle \equiv |n_1+i,n_2+j\rangle\, , \;\;\; i,j=0,1\, ,
\label{e2a} \end{equation}
and talk about the ``first'' and the ``second'' quasiparticle
on the antidots disregarding the background $n_l$ quasiparticles.
Counting the antidot energies $E_n$ from the energy of the state
with no extra quasiparticles, we can parameterize the energies
$\epsilon_{ij}$ of the four states as
\begin{equation}
\epsilon_{00}=0 \, , \;\;\; \epsilon_{11}= 2\epsilon + U \, , \;\;\;
\epsilon_{10}=\epsilon-\delta \, , \;\;\; \epsilon_{01}= \epsilon +
\delta \, . \label{e2}
\end{equation}
Here $U$ is the interaction energy between the extra quasiparticles
on the two antidots, and $\epsilon$ and $\delta$ give the energies
$\epsilon_l$ (Fig.~\ref{f1}b) of the single-quasiparticle states
localized at the two antidots: $\varepsilon_{1,2}= \epsilon \pm
\delta$. The energies $\epsilon$ and $\delta$ are defined
relative to the common chemical potential of the edges
for vanishing bias voltage $V$ between them. Non-vanishing bias
voltage shifts the energies (\ref{e2}). Experimentally, the antidot
energies are controlled by the back-gate voltage or magnetic field
\cite{ant1,ant2}. The degree to which these fields couple to the
energy difference $\delta$ depends on the degree of
asymmetry between the two antidots. In the following, we present the
results for quasiparticle conductance of the system as a function
of $\epsilon$ for fixed $\delta$, as would be appropriate for
identical antidots. These results can be generalized to non-identical
antidots by taking a ``cross-section'' in the space of $\epsilon$ and
$\delta$ along the direction appropriate for a given degree of the
antidot asymmetry.

If the two antidots are sufficiently close, so that the distance
between their edges is on the order of magnetic length $\ell$, the
quasiparticle states localized around them overlap and hybridize.
This effect can be accounted for by the tunnel coupling $-\Delta$
of the antidots. The phases of the antidot states can always be
chosen to make $\Delta$ real. This coupling affects only the
singly-occupied states $|10\rangle$ and $|01\rangle$. The
single-quasiparticle part of the Hamiltonian is then:
\begin{equation}
H = \epsilon -\delta \sigma_z -\Delta \sigma_x  \, , \label{e3}
\end{equation}
where $\sigma$'s are the Pauli matrices. Equation (\ref{e3}), together
with the part of Eq.~(\ref{e2}) describing states with zero and two
quasiparticles, give the main part of the antidot energy
controlling the quasiparticle transport. In what follows, we assume
that all contributions to this energy and the temperature
$T$ are small:
\begin{equation}
\Delta, \varepsilon_l, U, T \ll \Delta^*. \label{e4}
\end{equation}
In this regime, tunneling through the antidots can be discussed in
terms of correlated transfer of individual quasiparticles. Since the
gap $\Delta^*$ is dominated by the Coulomb interaction, i.e. has
the same origin as the quasiparticle interaction energy $U$, the
most restrictive part of the assumption (\ref{e4}) is the condition
on $U$. This condition can still be satisfied,
due to the difference between stronger quasiparticle repulsion on
the same site and weaker repulsion on different sites.

\subsection{Antidot-edge tunneling}
\label{tun}

Similarly to the tunnel coupling between the antidots, if the
edges of the FQHE liquid are not far from the antidots on the
scale of the magnetic length $\ell$, there is a non-vanishing
amplitude for quasiparticle tunneling between the edge and the
nearest antidot. The tunneling between the $l$th edge and
antidot can be described quantitatively with the standard
tunnel Hamiltonian
\begin{equation}
H_T^{(l)}= T_l \psi_l^{\dagger} \xi_l + h.c. \, , \label{e5}
\end{equation} where $\psi, \psi^{\dagger}$ and $\xi,
\xi^{\dagger}$ are the creation/ahhihalation operators for
quasiparticles at, respectively, the edges and the antidots.
Denoting the position along the edge as $x$ and taking the tunneling
points for both edges to be at $x=0$, the edge quasiparticle
operators $\psi_l$ can be expressed in the standard bosonisation
approach as \cite{wen}
\begin{equation}
\psi_l(t) =(1/2\pi \alpha)^{1/2} \tilde{\xi}_l e^{i\sqrt{\nu}
\phi_l(0,t)}\, . \label{e6} \end{equation} Here the ``Klein
factors'' $\tilde{\xi}_l$ account for the mutual statistics of
the quasiparticles in different edges, $\phi_l$ are the chiral
bosonic fields which describe the edge fluctuations propagating
with velocity $u$, and $1/\alpha$ is their momentum cut-off. The
edge fluctuations result in the fluctuations of electron density
at the edge: $\rho_l(x,t)=
(\sqrt{\nu}/2 \pi) \partial \phi_l(x, t)/\partial x$. Both fields
$\phi_l$ can be decomposed in the standard way into the individual
``magneto-plasmon'' oscillator modes $a_n, a_n^{\dagger}$:
\begin{equation}
\phi(x,t) = \sum_{n=1}^{\infty} \frac{1}{\sqrt{n}}\left[ a_n
e^{iq_n(x+i\alpha)} + h.c.\right]  , \;\; q_n=2\pi n/L \, ,
\label{e7}
\end{equation} where $L$ is a normalization length. In the geometry
of the double-antidot system, there are no interference loops for
the edge quasiparticle. In this case, the statistical Klein factors
in (\ref{e6}) would cancel out in the perturbation expansion in
tunneling (\ref{e5}) and can be omitted.

The quasiparticles at the antidots should be described in general by
the expressions similar to Eq.~(\ref{e6}). Condition (\ref{e4}) of
the large antidot energy gap $\Delta^*$ ensures, however, that the
fluctuations of the edges around the antidots are suppressed, i.e.
the magneto-plasmon oscillations are not excited out of their ground
state $|0\rangle$. In this regime of the ``quantized'' edge, the
general quasiparticle operators (\ref{e6}) reduce to just the
statistical Klein factors up to a normalization constant. Indeed, as
one can see directly from Eq.~(\ref{e7}) by bringing the $\phi$-part
of (\ref{e6}) into the normal form,
\begin{equation}
\langle 0| e^{i\sqrt{\nu} \phi}|0\rangle /(2\pi \alpha)^{1/2} = (\pi
R)^{-1/2} \, . \label{e8}
\end{equation}
Including this normalization constant in the tunnel amplitude
$T_l$, we see that the operators $\xi_l$ for quasiparticles at the
antidots consist solely of the Klein factors. The appropriate set
of properties of the quasiparticle Klein factors $\xi$ depends
on the specific geometry of each edge-state tunneling problem.
Non-trivial examples of this can be found in
\cite{kf1,kf2,kf4,we,kf3}. As follows from the discussion in the
preceding Section, in tunneling between the quantum antidots, the
operators $\xi_l$ should account for the ``hardcore'' property of
the quasiparticles: In the given range of external
parameters only one extra quasiparticle can occupy one antidot. In
general, these operators should also describe the anyonic exchange
statistics of the FQHE quasiparticles, but the geometry of the
double-antidot system (Fig.~\ref{f1}a) does not permit quasiparticle
exchanges, and the exchange statistics of the tunneling
particles is irrelevant. Since the hardcore property makes the
quasiparticle occupation factors equivalent to those of the
fermions, and the actual exchange statistics is irrelevant, the
antidot quasiparticle operators $\xi, \xi^{\dagger}$ can be
treated as fermions. Together with Eqs.~(\ref{e6}) and (\ref{e7})
for the edge quasiparticles, this defines completely the tunnel
Hamiltonian (\ref{e5}).

\subsection{Edge-state decoherence}
\label{dec}

Tunneling of charged quasiparticles through the antidot system
couples to all gapless charged excitations that exist in the system.
In the case of the FQHE liquid, excitations in the bulk of the liquid
are suppressed by the energy gap, and only the edges support gapless
excitations. In contrast to all other possible mechanisms of
decoherence (e.g., plasmons in metallic gates, or charged impurities
in the substrate) the edges play the role of reservoirs in transport
measurements and as a matter of principle can not be removed from the
antidots.  In this Section, we estimate the strength of this
unavoidable edge-state decoherence for quasiparticle tunneling through
the double-antidot system.

The spectrum of the gapless edge excitations of one edge consists of
magneto-plasmon oscillations (\ref{e7}) with the Hamiltonian:
\begin{equation}
H_0 =(hu/L) \sum_{n=1}^{\infty} n a_n^{\dagger}a_n \, . \label{e9}
\end{equation}
We assume that the antidot system is symmetric, and a
quasiparticle sitting on the first antidot creates a potential
$V_l(x)$ along the $l$th edge. The quasiparticle dynamics governed
by the Hamiltonian (\ref{e3}) is coupled then to the fluctuations
of electron densities $\rho_l(x)$ at the edges through the interaction
Hamiltonian
\begin{equation}
H_{int} =\frac{1}{2} \sigma_z e \int dx V(x) (\rho_1(x)-\rho_2(x))
\, , \label{e10}
\end{equation}
where $V(x)=V_1(x)-V_2(x)$ is the change of the potential along the
edges due to quasiparticle transfer between the antidots. Since the
edge-antidot distance and the distance between the antidots are on
the order of antidot radius $R$, this radius sets the range of the
potential $V(x)$. The edge velocity $u$ can be expected to be similar
for the external edges and the antidots. This means that the
condition (\ref{e4}) of the large energy gap implies that the
characteristic wavelength of the edge excitations which can exchange
energy with the quasiparticles on the antidots is much larger than
the range of the potential $V(x)$: $\hbar u / \epsilon \gg \hbar u /
\Delta^* \simeq R$. The interaction energy (\ref{e10}) can be
expressed then as
\begin{equation}
H_{int} =  \frac{e}{2} \sigma_z (\rho_1(0)-\rho_2(0)) \int dx V(x)
\, , \label{e11} \end{equation} where, as follows from
Eq.~(\ref{e7}), the densities $\rho_l$ are
\begin{equation}
\rho_l (0) = \frac{i\sqrt{\nu} }{L} \sum_{n=1}^{\infty} \sqrt{n} (a_n-
a_n^{\dagger} ) \, . \label{e12} \end{equation}

The strength of interaction (\ref{e11}) can be characterized by
the typical transition rate $\Gamma_d$ between the eigenstates of
the antidot Hamiltonian (\ref{e3}) induced by the
edges. Straightforward calculation of the ``Golden-rule'' rate
using Eqs.~(\ref{e9}), (\ref{e11}), and (\ref{e12}) gives:
\begin{equation}
\Gamma_d =  \frac{\nu^3}{4 \pi \hbar} \alpha^2 \kappa^2 |\langle
\sigma_z \rangle|^2 \frac{\Delta E}{1-e^{-\Delta E/T}} \, ,
\label{e13} \end{equation} where $\Delta E$ is the energy difference
between the two states, $\langle \sigma_z \rangle$ is the matrix
element of $\sigma_z$ between them. The dimensionless
factor $\kappa$ characterizes the overall ``intensity'' of the
antidot-edge potential:
\begin{equation}
\kappa \equiv \left( \frac{ \nu e }{4\pi \varepsilon
\varepsilon_0} \right)^{-1} \int dx V(x) \, , \label{e14}
\end{equation} The precise form of the potential $V(x)$ and the
value of $\kappa$ depend on the details of configuration of the
metallic gates that define the edges and screen the antidot-edge
interaction. However, normalized as in Eq.~(\ref{e14}), $\kappa$
should be on the order of 1. For instance, assuming as a crude
model of the system electrostatics that the antidot-edge
interaction is confined to the interval $d\simeq 2R$ in which the
edge is a tunnel-limited distance $\ell$ away from the antidot,
one can estimate $\kappa$ as $(2/\pi )\ln(2R/\ell)$, i.e., $\kappa
\simeq 2$ for realistic $R/\ell \simeq 10$.

The factor $\alpha$ in (\ref{e13}) is the ``fine structure
constant'' of the edge excitations:
\begin{equation}
\alpha \equiv \frac{e^2}{4\pi \varepsilon \varepsilon_0 \hbar u}
\, , \label{e15} \end{equation} and is the main parameter
controlling the strength of decoherence $\Gamma_d$ through the
velocity $u$ of the edge excitations. The dielectric constant
$\varepsilon$ is fixed by the material (GaAs) of the structure:
$\varepsilon \simeq 10$, and in the realistic range of possible
velocities $u$, $10^4 \div 10^5$ m/s \cite{ant2}, $\alpha$ should
vary in the range between 2 and 20. In the most relevant case of
the FQHE liquid with the filling factor $\nu=1/3$, and for the
edge-antidot coupling intensity estimated above, the quality
factor $\Delta E/\hbar \Gamma_d$ of the quasiparticle dynamics
changes then roughly between 0.1 and 10. This means that in the
case of strong edge confinement that produces large velocity $u$,
the quasiparticle dynamics on the antidots can be quantum-coherent
provided that all other decoherence mechanisms are sufficiently weak.
In the opposite case of smooth confinement with low velocity $u$,
the already unavoidable edge-state decoherence is strong enough
to completely suppress the coherence of the quasiparticle states
on different antidots, and quasiparticle transfer processes between
them are incoherent.

\section{Tunneling rates}

As was mentioned above, the discussion in this work is limited to
the regime in which the transport through the double-antidot
system can be interpreted as the correlated transfer of individual
quasiparticles. Besides the condition (\ref{e4}) on antidot
energies, this also requires that the antidots are coupled only
weakly to the edges, so that the edge-antidot tunneling can be
treated as a perturbation leading to an incoherent transfer of
individual quasiparticles. The quasiparticle transport through the
antidots is governed then by the kinetic equation similar to that
for Coulomb-blockade transport in quantum dots with discrete
energy spectrum \cite{b8}. This Section calculates the relevant
tunneling rates in the two limits of strong and weak edge-state
decoherence.

\subsection{Strong decoherence}

\begin{figure}[h]
\begin{center}
\includegraphics[scale=.31]{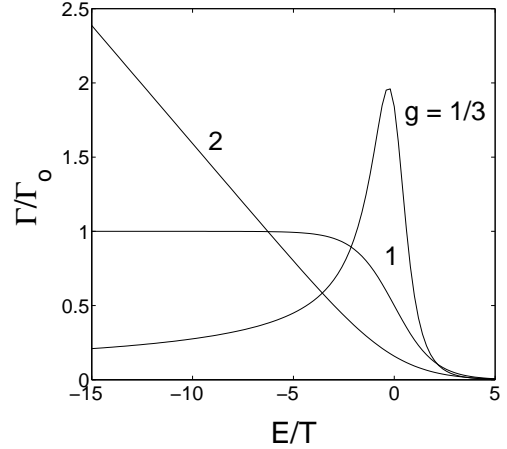}
\end{center}
\setlength{\unitlength}{1.0in}
\caption{Energy dependence of the antidot-antidot \protect (\ref{e19})
and antidot-edge (\ref{e20}) tunneling rates. The normalization factor is
$\Gamma_0= \gamma (2\pi T/\omega_c)^{g-1}$. }
\label{f2} \end{figure}

If the edge-state decoherence is sufficiently strong, the
quasiparticle transfer between the antidots can be treated
as incoherent and described by the sequential tunneling rate obtained
by perturbation theory in the tunnel amplitude $\Delta$. To calculate
this rate, it is convenient to express the density operators $\rho_l$
of the two edges through one effective density $\rho$ which satisfies
the same relation (\ref{e12}):  $\rho_1(0)-\rho_2(0) \rightarrow
\sqrt{2} \rho(0)$, so that the edge-antidot coupling (\ref{e11}) is:
\begin{equation}
H_{int} =  \hbar u \, \nu \frac{\kappa \alpha }{\sqrt{2}}\, \rho(0)
\, \sigma_z \, . \label{e16} \end{equation} Next, one can perform
a unitary transformation which converts the fluctuations of the
energy of the quasiparticle basis states $|10\rangle \,
,|01\rangle$ induced by (\ref{e16}) into a fluctuating phase of
the tunneling matrix elements of the tunneling part of the
Hamiltonian (\ref{e3}):
\begin{equation}
- \Delta \sigma_x \;\; \rightarrow - \Delta \sum_{\pm} \sigma_{\pm}
e^{\pm i \sqrt{g} \phi(0,t)},\;\;\;\; g= \frac{\nu^3 \kappa^2 \alpha^2
}{2\pi^2} \, , \label{e17} \end{equation}
where $\phi(x,t)$ is the bosonic field given by same the Eq.~(\ref{e7}).
Then, the rate
$\Gamma_{\Delta}$ of the {\em anidot-antidot} tunneling can be expressed
in the lowest non-vanishing order in the amplitude $\Delta$ as:
\begin{equation}
\Gamma_{\Delta} = 2 \Delta^2 \mbox{Re} \int_{-\infty}^{0} dt e^{i Et}
\langle e^{i \sqrt{g} \phi(0,t)}e^{-i \sqrt{g} \phi(0,0)} \rangle \, ,
\label{e18} \end{equation}
where $\langle ... \rangle$ is the average over the equilibrium fluctuations
of $\phi$ and $E=\pm 2\delta$ is the energy difference (depending on the
direction of tunneling) between the quasiparticle states localized on the
antidots. The standard evaluation of Eq.~(\ref{e18}) (see, e.g., \cite{b6,b7})
gives:
\begin{equation}
\Gamma_{\Delta} (E) = \gamma f_g(E) \, , \;\;\; \gamma \equiv
2\pi \Delta^2/\omega_c \, , \label{e19} \end{equation}
\[f_g(E) \equiv \frac{1}{2\pi \Gamma(g)}(2\pi T/\omega_c)^{g-1} \left| \Gamma
(g/2 + i E/2\pi T )\right|^2 e^{-E/2T} \, , \] where $\Gamma(z)$
is the gamma-function and $\omega_c = \hbar u/2\alpha$ is the
cut-off energy of the edge excitations. The function $f_g(E)$
gives the energy dependence of the tunneling rate (see
Fig.~\ref{f2}) and is defined to coincide with the Fermi
distribution function for $g=1$. The power $g$ determines the
behavior of the transition rate at large energies $|E|\gg T$:
$\Gamma_{\Delta} (E) \propto E^{(g-1)}$ on the ``allowed'' side of
the transition ($E<0$), and $\Gamma_{\Delta} (E) \propto
E^{(g-1)}e^{-E/T}$ on the ``forbidden'' side ($E>0$), when the
transition has to overcome the energy barrier $E$. This asymptotic
behavior of the tunneling rates, together with Eq.~(\ref{e19}), is
valid at $|E| \ll \omega_c$.

The rates $\Gamma_l$, $l=1,2$, of the {\em antidot-edge} tunneling
are obtained through a similar calculation starting with the
tunnel Hamiltonian (\ref{e5}). They are given by the same
expression (\ref{e19}):
\begin{equation}
\Gamma_l (E) = \gamma_l f_{\nu} (E) \, , \;\;\; \gamma_l \equiv
2\pi |T_l|^2/\omega_c\, . \label{e20} \end{equation}
In general, the long-range Coulomb interaction should generate corrections
to $g$ which move it away from the ``quantized'' value $g=\nu$ \cite{b9}.
However, in contrast to the quasiparticle tunneling between the antidots,
which is changed qualitatively by decoherence created by the Coulomb
interaction with the edge, the Coulomb corrections for the antidot-edge
tunneling are expected to be small and will be neglected in this work.

Thus, in the regime of strong edge-state decoherence, the overall
transport of quasiparticles through the double-antidot system can
be described as a combination of successive antidot-edge
transitions (\ref{e20}) and incoherent transitions (\ref{e19})
between the antidots.

\subsection{Weak decoherence}

For sufficiently strong edge-state confinement, the edge-induced
relaxation rate (\ref{e13}) will be smaller than the antidot
energies. If other decoherence mechanisms, including the decoherence
created by incoherent antidot-edge tunneling are also weak on the
scale of the antidot energies,
\begin{equation}
\Gamma_d, \Gamma_l \ll \Delta, \delta, U, T, \label{e21}
\end{equation}
the quasiparticle dynamics on the antidots is
quantum-coherent. It is characterized by the stationary
eigenstates $|k\rangle $, $k=1,2$, of the double-antidot
Hamiltonian (\ref{e3}):
\[ H |k\rangle  = (\epsilon +(-1)^k \Omega) |k \rangle ,
\;\; \Omega \equiv (\delta^2 +\Delta^2)^{1/2}, \]
\begin{equation}
|k\rangle =c_{1k}|10\rangle + c_{2k} |01 \rangle , \label{e22}
\end{equation} for which the probabilities $\lambda_{lk}$ of
finding the quasiparticle on the $l$th antidot are:
\begin{equation}
\lambda_{lk}=|c_{lk}|^2 = [1+ (-1)^{l+k}\delta/\Omega]/2 \, .
\label{e23}
\end{equation}

In the coherent regime (\ref{e21}), the double-antidot system can
be viewed as a quasiparticle qubit \cite{qub}. The current through
the qubit is described in terms of tunneling to/from the
eigenstates (\ref{e22}). The corresponding tunneling rates are
found from the tunnel Hamiltonian (\ref{e5}), in which, as was
discussed in Sec.~\ref{tun}, the quasiparticle
creation/annihalation operators $\xi, \xi^{\dagger}$ act as
fermions. This means that the tunnel matrix elements for the
quasiparticles can be calculated in the standard way. In
particular, for each eigenstate (\ref{e22}), the matrix element is
independent of the occupation factor of the other eigenstate.
Explicitly, the tunneling rate $\Gamma_{lk}$ from the $l$th edge
into the state $|k\rangle$ is
\begin{equation}
\Gamma_{lk} = 2 |T_l|^2 |\langle k| \xi_l^{\dagger}|0 \rangle |^2
\mbox{Re} \int_{-\infty}^{0} dt e^{i Et} \langle \psi_l^{\dagger}(t)
\psi_l (0) \rangle \, . \label{e51} \end{equation}
Here $|0 \rangle$ denotes the empty eigenstate and $E$ is the
appropriate tunneling energy which includes in general the eigenenergies
(\ref{e22}) and the interaction energy $U$. The quasiparticle matrix
elements are $ |\langle k| \xi_l^{\dagger} |0 \rangle |^2 = \lambda_{lk}$,
and we get:
\begin{equation}
\Gamma_{lk} (E) = \lambda_{lk} \Gamma_l (E) , \label{e24}
\end{equation}
where the rates $\Gamma_l (E)$ are given by Eq.~(\ref{e20}).
In the practically important case of FQHE liquid with $\nu=1/3$,
the energy dependence of the transition rates (\ref{e24}) is
illustrated by the $g=1/3$ curve in Fig.~\ref{f2}. The peak of
the tunneling rate at $\epsilon \simeq 0$ is the consequence
of the Luttinger-liquid correlations of the edge quasiparticles.
Conductance calculations presented in the next Section show that
this peak manifests itself as additional resonant features
of the qubit conductance.

\section{Conductance of the double-antidot system}

In both situations of strong and weak decoherence, the conductance
associated with tunneling of individual quasiparticles through the
double-antidot system can be calculated by solving the kinetic
equation for quasiparticle occupation probabilities of the antidot
states. Similarly to the case of tunneling through one antidot
\cite{ant1,ant2,ant3,gel}, the conductance as a function of the
common energy $\epsilon$ of the antidot states should exhibit the
resonant tunneling peaks. For the double-antidot system, the
peak structure is, however, more complicated, reflecting the transition
between the low-temperature regime in which each peak corresponds
to addition of one quasiparticle to the system of antidots, and a
possible ``large-temperature'' regime, when the single-quasiparticle
peaks are merged, and each conductance peak is associated with
addition of two
quasiparticles. In this Section, we calculate the corresponding
conductance line shapes. Quantitatively, these line shapes are
determined by the interplay between the quasiparticle repulsion
energy $U$ on the two antidots and tunnel coupling $\Delta$
between them. The calculations below are focused mostly on the
more typical case of large repulsion energy $U\gg \Delta \simeq
\delta$.

\subsection{Strong decoherence}

\begin{figure}[h]
\begin{center}
\includegraphics[scale=.32]{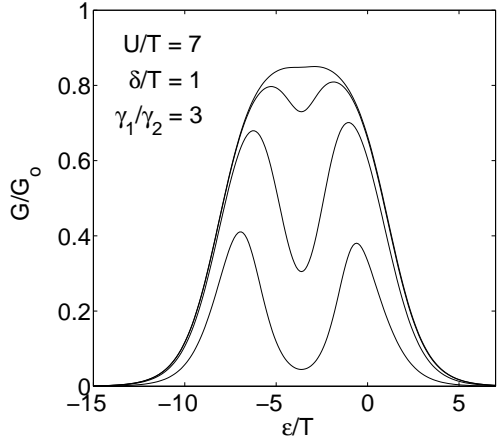}
\end{center}
\caption{Conductance of the double-antidot system in the regime of
overdamped quasiparticle transport. Conductance is normalized to
$G_0=(e\nu)^2\Gamma_{\Delta}(2\delta)/ T$. Different curves
correspond to different ratios of the antidot-antidot and
antidot-edge tunneling rates. From top to bottom:
$\Gamma_{\Delta}(2\delta)/\Gamma_2(0)= 10^{-4},10^{-3},10^{-2},
10^{-1}$.} \label{f3}  \end{figure}

For strong edge-state decoherence, coherent mixing of the
quasiparticle states on the two antidots is suppressed, and the
quasiparticle dynamics is described by kinetic equations for
the occupation probabilities $p_{ij}$ of the states $|ij\rangle$
(\ref{e2a}) localized on the antidots. The probabilities
evolve due to incoherent jumps of quasiparticles at the rates
$\Gamma_{\Delta}$ (\ref{e19}) and $\Gamma_l$ (\ref{e20}) between
these states. The stationary quasiparticle current $I$ through
the antidots is found in this regime from the balance of the
forward/backward transition across any of the three tunnel
junctions of the system, e.g., from the transitions between the
antidots:
\begin{equation}
I =e\nu [p_{10} \Gamma_{\Delta} (2\delta) - p_{01} \Gamma_{\Delta}
(-2\delta)]\, . \label{e26} \end{equation}
In general, the quasiparticle current
$I$ can be calculated by the direct numerical solution of the
kinetic equation. The results of such solution for the linear
conductance $G=dI/dV|_{V=0}$ are shown in Fig.~\ref{f3} (in all
numerical results presented below we take $\nu=1/3$). Qualitative
behavior of the system can be understood by analyzing the limits
where the simple analytical expressions for the conductance
can be obtained.

The first limit is $\Gamma_{\Delta}\ll \Gamma_l$, where the
antidot-antidot tunneling is the bottleneck for the current flow.
In this case, to the
zeroth-order approximation in $\Gamma_{\Delta}$, the current
is vanishing, and one can use in Eq.~(\ref{e26}) the equilibrium
probabilities $p_{ij}=(1/Z)e^{-\epsilon_{ij}/T}$, $Z=\sum_{ij}
e^{-\epsilon_{ij}/T}$, obtaining for the conductance:
\begin{equation}
G=\frac{(e\nu)^2\gamma}{T}\frac{f_g(2\delta)e^{\delta/2T}}{
e^{-\epsilon/T}+ e^{-(\epsilon+U)/T} +2\cosh(\delta/T)} \, .
\label{e27} \end{equation}

Equation (\ref{e27}) describes the ``coalesced'' conductance peak
that corresponds to the addition of two quasiparticles to the
antidots. At large temperatures, $T\geq U$, the peak has a
usual thermally-broadened shape with width proportional to $T$. At
$T\ll U$, however, the peak shape
(\ref{e27}) is quite unusual: the conductance is constant between
the point $\epsilon \simeq 0$, when the first quasiparticle is
added to the antidots, and the point $\epsilon \simeq -U$ of
addition of the second quasiparticle, forming the plateau of width
$U$ -- see Fig.~\ref{f3}. The conductance plateau remains flat
until the temperature is lowered to $T\simeq U/\ln [\Gamma_l
(U)/\Gamma_{\Delta}]$, when the thermal suppression of the
antdot-edge tunneling rate makes it comparable to
$\Gamma_{\Delta}$ at the center of the plateau, $\epsilon \simeq
-U/2$, despite the fact that the two rates are very different at
$\epsilon \simeq 0$. In this temperature range, a dip develops in
the center, which separates the plateau into two peaks, one at
$\epsilon \simeq 0$ and the other at $\epsilon \simeq -U$, with
decreasing temperature (Fig.~\ref{f3}). Each peak corresponds to
addition of one quasiparticle to the double-antidot system.
Note that the resonant peaks occur when the gate bias energy
$\epsilon$ is equal to ``minus energy'' of the antidot
state, so that the total energy of the state relative to the
chemical potential of the edges is zero.

The shape of such single-quasiparticle peaks can be described
in the opposite limit of strong
antidot-antidot tunneling $\Gamma_{\Delta} \gg \Gamma_l$. In this
limit, the general kinetic equations for three probabilities
$p_{00}, p_{10}, p_{01}$ relevant at $\epsilon \simeq 0$, e.g.,
\begin{equation}
\dot{p}_{00}=\Gamma_1(-\epsilon_1)p_{10}+\Gamma_2(-\epsilon_2)p_{01}
-[\Gamma_1 (\epsilon_1)+\Gamma_2 (\epsilon_2)]p_{00}, \label{e25}
\end{equation} and similar equations for the other probabilities,
can be reduced to two equations for the effective two-state
system. The strong antidot-antdot tunneling that couples the
singly-occupied states $|10\rangle\, ,|01\rangle$, maintains the
relative equilibrium between them: $p_{10}/p_{01}=e^{-2\delta/T}$,
making it possible to treat these two states as one. The effective
transition rates between this state and the state $|00\rangle$ are
obtained as the weighted average of the transition rates in
starting kinetic equations, e.g. (\ref{e25}). The standard
calculation of the current through a two-state system gives then
the resonant peak of the double-antidot conductance at $\epsilon
\simeq 0$ (associated with addition of the first quasiparticle to
the antidots):
\begin{equation}
G=\frac{(e\nu)^2 \Gamma_1(\epsilon_1)\Gamma_2 (-\epsilon_2)
f(2\delta)/T }{\Gamma_1(\epsilon_1)+ \Gamma_2 (\epsilon_2)+
\Gamma_1(-\epsilon_1)f(-2\delta)+
\Gamma_2(-\epsilon_2)f(2\delta)}\, , \label{e29} \end{equation}
where $f(E)=f_1(E)$ is the Fermi distribution function. For
instance, if the two antidot states are aligned, $\delta=0$,
\begin{equation}
G=\frac{(e\nu)^2}{T} \frac{\gamma_1 \gamma_2}{\gamma_1+ \gamma_2}
\frac{f_{\nu} (\epsilon)}{1+2e^{-\epsilon/T}}\, .
\label{e30} \end{equation}

The conductance peak at $\epsilon \simeq -U$ associated with the
addition of the second quasiparticle is given by an expression
similar to Eq.~(\ref{e29}) with an appropriate shift of energy
$\epsilon \rightarrow \epsilon +U$. In particular, for $\delta=0$,
this expression reduces to
\begin{equation}
G=\frac{(e\nu)^2}{T} \frac{\gamma_1 \gamma_2}{\gamma_1+ \gamma_2}
\frac{f_{\nu} (\epsilon +U)}{2+e^{-(\epsilon +U)/T}}\, . \label{e31}
\end{equation} The conductance peak (\ref{e30}) is asymmetric around
$\epsilon =0$, since in the tunneling dynamics underlying this
peak, only one quasiparticle can tunnel off the antidots, while
there are two available states for tunneling onto the antidots.
Still, the ``quasiparticle-quasihole'' symmetry makes the two
peaks, (\ref{e30}) and (\ref{e31}), at $\delta=0$ symmetric images
of each other with respect to a ``mirror'' reflection
$\epsilon+U/2 \rightarrow -(\epsilon +U/2)$. For $\delta \neq 0$,
the condition $\Gamma_{\Delta} \gg \Gamma_l$ is violated at
sufficiently low temperatures $T\ll \delta$, and Eq.~(\ref{e29})
becomes invalid. In this case, the antidots are effectively out of
resonance, and conductance peaks are suppressed exponentially with
temperature at all gate bias energies $\epsilon$.

\subsection{Weak decoherence}

If the edge-state decoherence is sufficiently weak and allows for
quantum-coherent transfer of quasiparticles between the two
antidots, the kinetic equation for quasiparticle transport should
be written not in the basis of states, (\ref{e2a}), but in the
basis of the hybridized states (\ref{e22}). As follows from the
estimates of the edge-state decoherence in Sec.~\ref{dec}, even in
this regime, the edge-induced relaxation rate $\Gamma_d$
(\ref{e13}) should be strong enough, $\Gamma_d \gg\Gamma_l$, to
maintain the equilibrium distribution of quasiparticles over the
antidot states in the process of tunneling. This means that if
$E_k^{(n)}$ is the energy of the state $|k\rangle$ when there are
$n$ quasiparticles on the antidots, the probability that this
state is occupied is $\rho_k(n) = (1/Z_n)e^{-E_k^{(n)}/T }$,
$Z_n=\sum_ke^{-E_k^{(n)}/T }$. The quasiparticle tunneling is
reduced then to the dynamics of the total number $n$ of
quasiparticles on the antidots, described by the probability
distribution $p(n)$. The rates of tunneling transitions
$n\rightarrow n\pm 1$ in this dynamics are:
\[ \Gamma^{\pm}(n)=\sum_{l=1,2}\Gamma^{\pm}_l (n),\;\;
\Gamma^{\pm}_l (n) = \sum_{kq}\rho_k(n) \Gamma_l (k,q,n,n\pm 1)\,
. \]
where the partial transitions rates
$\Gamma_l (p,k,n,n\pm 1)$ from the state $k$ of $n$ quasiparticles
into the state $k'$ of $n\pm 1$ quasiparticles are given by the
appropriate tunneling rates (\ref{e24}) between the $l$th edge and
antidot. The solution of the simple kinetic equation
\begin{equation}
\dot{p}(n) = \sum_{\pm}\left[\Gamma^{\mp}(n\pm1)p(n\pm1)
-\Gamma^{\pm}(n)p(n)\right] \, , \label{e36}
\end{equation}
gives then the stationary quasiparticle current through
the system:
\begin{equation}
I = \nu e \sum_n \left[\Gamma^{+}_1(n)P(n) -
\Gamma^{-}_1(n-1)P(n-1) \right]\, . \label{e37}
\end{equation}

\begin{figure}[h]
\begin{center}
\includegraphics[scale =.30]{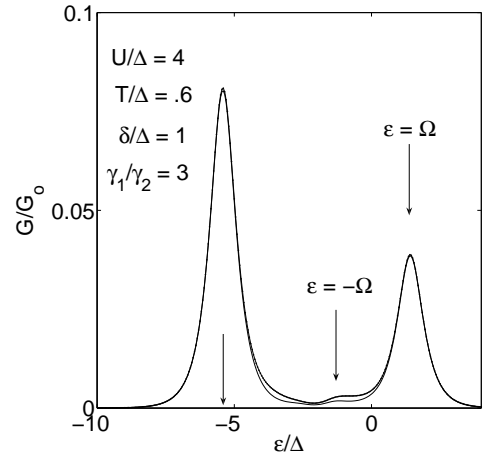}
\end{center}
\caption{Conductance of the double-antidot system in the regime of the
underdamped quasiparticle dynamics. Conductance is plotted in units
of $G_0=(e\nu)^2\Gamma_1(0)/T$. The curves show the two main resonant
conductance peaks at $\epsilon=\Omega$ and $\epsilon=-(U+\Omega)$,
and a weak kink at $\epsilon=-\Omega$ that is made visible by the
Luttinger-liquid singularity in the tunneling rates. The upper
and lower curves are, respectively, the conductance with and without
equilibration on the antidots. }
\label{f4} \end{figure}

Equation (\ref{e36}) shows that the stationary probability
distribution $p(n)$ satisfies the ``detailed balance'' condition
$p(n)\Gamma^{+}(n) = p(n+1)\Gamma^{-}(n+1)$ even in the presence
of the non-vanishing bias voltage $V$. Using this condition, and
expanding both $p(n)$ and the tunneling rates $\Gamma(n)$ to first
order in $V$, one finds the linear conductance $G$ of the
quasiparticle qubit:
\begin{equation}
G =\eta \frac{(e\nu)^2}{T}\sum_n w_n \frac{\Gamma^{+}_1(n)
\Gamma^{+}_2(n) }{\Gamma^{+}_1(n)+\Gamma^{+}_2(n)}\, . \label{e38}
\end{equation}
Here $w_n=Z_n/Z$ is the equilibrium probability to have $n$
quasiparticles on the antidots, $Z= \sum_n Z_n$, and
the factor $\eta$ gives the fraction of the voltage $V$ that drops
across the edge-antidot junctions. Equation (\ref{e38}) can be
understood in terms of forward jumps of quasiparticles in the left
junction contributing to the current only if they are followed by the
forward jumps in the right junction. As an example, at temperatures
$T\ll U$, and $\epsilon \simeq -\Omega$ one can limit the sum in
Eq.~(\ref{e38}) to one term $n=0$. The conductance $G$ is then:
\[ G =\frac{(e\nu)^2}{T}\frac{\eta }{1+2 e^{-\epsilon /T} \cosh
(\Omega /T)}\,  \cdot \]
\begin{equation}
\frac{\sum_{q,k} \Gamma_{1q}(\epsilon +(-1)^q\Omega) \Gamma_{2k}
(\epsilon +(-1)^k\Omega)}{\sum_{l,m}\Gamma_{lm}(\epsilon +
(-1)^m\Omega)}\, , \label{e39} \end{equation} where the tunneling
rates $\Gamma_{qk}$ are defined in Eq.~(\ref{e24}).

\begin{figure}[h]
\hspace*{0.2in}
\includegraphics[scale=.29]{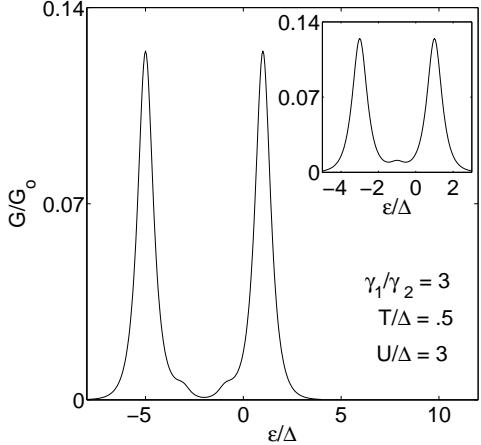}
\caption{Conductance of the symmetric ($\delta=0$) antidot qubit
exhibiting two resonant conductance peaks at $\epsilon= \Delta$
and $\epsilon=-(U+\Delta)$. Both peaks have kinks at $\epsilon=
-\Delta$ and $\epsilon=-(U-\Delta)$ caused by the Luttinger-liquid
singularity in the tunneling rates. The inset shows the conductance
for the special value of interaction energy $U=2\Delta$, when the
two kinks coincide producing very small but visible additional
conductance peak. Conductance is normalized as in \protect
Fig.~\ref{f4}. }
\label{f5}  \end{figure}

For comparison, one can calculate the conductance in the same
regime $T\ll U$, $\epsilon \simeq -\Omega$, but without
equilibration on the antidots, i.e. assuming that the edge-state
decoherence is very weak, $\Gamma_d \ll \Gamma_l$. As before, the
antidots can be occupied in this regime at most by one
quasiparticle at a time, and straightforward solution of the
kinetic equation describing the occupation of individual energy
eigenstates due to transitions (\ref{e24}) gives the conductance:
\[ G =\frac{(e\nu)^2}{T}\frac{\Delta^2}{2\Omega^2} \frac{\eta
\gamma_1 \gamma_2 }{1+ 2 e^{-\epsilon /T} \cosh (\Omega /T)} \,
\cdot \]
\begin{equation}
\sum_{\pm } \frac{f_{\nu}  (\epsilon \pm \Omega)}{\gamma_1 (1 \mp
\delta/\Omega) + \gamma_2 (1 \pm \delta/\Omega)}\, . \label{e40}
\end{equation} Equation (\ref{e40}) describes the resonant
conductance peak that corresponds to the addition of the first
quasiparticle to the antidot. The second quasiparticle peak at
$\epsilon = -(U+\Omega)$ is described by the similar expression.
At low temperatures, $T\ll \Omega$, only the lowest energy
eigenstate with energy $-\Omega$ contributes to the conductance
(\ref{e40}). In this case, the equilibration on the antidots does
not have any effect, and Eqs.~(\ref{e39}) and (\ref{e40})
coincide. As one can see from Fig.~\ref{f4}, which plots the
conductance obtained by numerical solution of the full kinetic
equation, the difference between the two regimes, with and without
relaxation, remains very small even at moderate temperatures. At
larger temperatures, $\Omega \ll T \ll U$, and $\delta=0$,
Eq.~(\ref{e40}) reduces to Eq.~(\ref{e30}) for the conductance in
the overdamped regime. The only difference between the two results
is the factor $\eta$ in Eq.~(\ref{e40}) which implies that the
part of the applied bias voltage that drops across the region of
the quantum-coherent quasiparticle dynamics does not contribute to
the linear conductance.

Besides the two main resonant peaks, the curves in Fig.~\ref{f4}
exhibit also a small kink at $\epsilon \simeq - \Omega$. This
kink appears at the intermediate temperatures and is the result
of the transfer of the first quasiparticle added to the antidots
not through the more probable ground state of the qubit but
through the excited state with energy $\Omega$. One could see,
however, by plotting the conductance of the double-antidot
system for tunneling electrons (the tunneling rates given by
the $g=1$ in Fig.~\ref{f2}) that the contribution of the excited
state to the conductance is not sufficient by itself to produce
such a kink. The kink in the conductance appears only
when the contribution from the excited state is amplified by the
Luttinger-liquid singularity in the quasiparticle tunneling rate
(seen in the $g=1/3$ curve in Fig.~\ref{f2} as a peak at
zero energy). It becomes somewhat more pronounced in the
conductance peaks of the ``symmetric'' qubit with $\delta=0$
shown in Fig.~\ref{f5}. In this case, the kinks appear on both peaks:
at $\epsilon=-\Delta$ and $\epsilon=-(U-\Delta)$. The second kink
is due to transport through the ground state of the qubit in the
regime when the main contribution to conductance comes from
the excited state. As shown in the inset in Fig.~\ref{f5},
at the special value of the interaction energy $U\simeq 2\Delta$,
the two kinks coincide and form a weak additional peak of the
qubit conductance.

\section{Conclusion}

We have calculated the linear conductance $G$ of
the double-antidot system in the regime of weak quasiparticle
tunneling through the antidots. Depending on the strength of the
edge-state decoherence, the tunneling can be coherent or
incoherent. In the incoherent regime, the two resonant
conductance peaks that correspond to the two antidot states are
spaced by the quasiparticle interaction energy $U$. In the
coherent regime, this spacing is increased to $U+2\Omega$, where
$2\Omega$ is the gap between the energy eigenstates of the
double-antidot system. The coherent regime of quasiparticle
dynamics is also characterized by the Lorentzian dependence
of the system conductance, $G\propto (1+\delta^2/\Delta^2)^{-1}$,
on the energy difference $\delta$ between the antidots.
In the quantum-coherent regime, the double-antidot system can
be used as a quasiparticle qubit.

The authors would like to thank V.J. Goldman for many useful
discussions of the antidot transport. This work was supported
in part by NSF grant \# DMR-0325551 and by ARO grant \#
DAAD19-03-1-0126.

\end{document}